\documentstyle[12pt]{article}

\newcommand{\be}{\begin{equation}}
\newcommand{\ee}{\end{equation}}
\newcommand{\bea}{\begin{eqnarray}}
\newcommand{\eea}{\end{eqnarray}}

\newcommand{\V} {{\cal V}_{n}}

\begin{document}

\begin{center}
\begin{large}
{\bf   AdS Holography     \\}
{\bf   and  \\}
{\bf   Strings on the Horizon  \\}
\end{large}  
\end{center}
\vspace*{0.50cm}
\begin{center}
{\sl by\\}
\vspace*{1.00cm}
{\bf A.J.M. Medved\\}
\vspace*{1.00cm}
{\sl
Department of Physics and Theoretical Physics Institute\\
University of Alberta\\
Edmonton, Canada T6G-2J1\\
{[e-mail: amedved@phys.ualberta.ca]}}\\
\end{center}
\bigskip\noindent
\begin{center}
\begin{large}
{\bf
ABSTRACT
}
\end{large}
\end{center}
\vspace*{0.50cm}
\par
\noindent
The microscopic origin of black hole entropy remains
one of the more intriguing open questions in theoretical
physics. A subplot in this drama is the renowned  Cardy-Verlinde
formula, which uses  two-dimensional conformal formalism
to explain  the  entropy of  an arbitrary-dimensional
black hole. In this paper, by exploiting the AdS/CFT and black hole-string
dualities, we are able to provide a physical picture
for this paradoxical behavior.
 Following a recent study
by Halyo (in a dS context), we  show that the dual CFT for
an asymptotically AdS spacetime actually conforms to
a string-like description. Moreover, we  demonstrate
that this stringy CFT is  directly
related  
to a string that lives on the stretched horizon
of an AdS-Schwarzschild-like  black hole.
In fact, after an appropriate renormalization, these two boundary theories
are shown to be thermodynamically equivalent.

\newpage

\section{Introduction}

One of the major tests for any prospective theory of quantum gravity
will be its ability to explain the Bekenstein-Hawking (black
hole) entropy \cite{bek,haw} from a microscopic perspective.
That is to say, the definitive statistical-mechanical explanation
for this entropy, which has its origins in purely thermodynamic
principles, is still conspicuously lacking. 
For a general overview on this puzzlement and  various  attempts
at its resolution, we refer the reader to Ref.\cite{crev}.
It is worth noting, however, that many such attempts
have, in fact, reproduced the anticipated Bekenstein-Hawking form.
The implied ``failure'' has, rather, been at the interpretative level;
in particular, a lack of clarity as to what degrees of freedom
are being counted from the perspective of a black hole.
\par
Significantly,
many state-counting formulations of black hole
entropy have utilized the well-known Cardy formula \cite{car}
as a key, intermediate step.  In fact, some  interesting  treatments
by Carlip \cite{carlip} and Solodukhin \cite{sol} have
shown that the Cardy formula is relevant  to the entropic
calculation
of   any  theory that admits  a black hole horizon. (Although
there may be some yet-unknown pathological exceptions.)
What is both intriguing and, at the same time, puzzling is
that, strictly speaking, the Cardy formula counts the degrees of freedom
in a two-dimensional conformal field theory (CFT). Paradoxically,
the cited methods apply to virtually any black  hole of 
arbitrary dimensionality.
That is to say, the black hole horizon should, naively, only
be expected to host a two-dimensional effective theory in the special case
of a three-dimensional spacetime.
 This logic follows from the holographic
principle  \cite{tho,sus}, which implies that a $d$+1-dimensional
bulk theory can  effectively be described by a $d$-dimensional
boundary theory.\footnote{Alternatively, this logic follows from
 a commonly accepted viewpoint: the degrees of freedom of a black
hole will predominantly live at or near the horizon \cite{bw}.}
\par
Interestingly, the Cardy formula  plays a significant role
in  yet another facet of black hole entropy.  Not too long
ago, Verlinde \cite{ver},   exploiting the well-known  duality
between anti-de Sitter (AdS) spacetimes and CFT's \cite{mal,gub,wit},
was able to show that the   entropy of an 
 AdS-Schwarzschild black hole
can be expressed in a Cardy-like form. That is,  given such a $d$+1-dimensional
 asymptotically AdS spacetime, the  thermodynamic
properties of an appropriately dual ($d$-dimensional) CFT   can be arranged
into a replica of the Cardy formula;  provided that a suitable
identification with the Cardy ``central charge'' \cite{car} has been made.
It just so happens that this Cardy-like entropy of the   CFT
is identical to that of the AdS-black hole bulk \cite{wit}.
(Note that the dual nature of such a CFT, which
 lives on a $d$-dimensional timelike boundary of the AdS bulk,
 follows from
the  AdS/CFT correspondence \cite{mal,gub,wit}. This correspondence, in turn,
can be viewed as a direct consequence of the holographic principle
\cite{tho,sus}.) 
\par
Verlinde's particular formulation \cite{ver} of the 
AdS/CFT entropy is now commonly referred to
as the Cardy-Verlinde formula. This basic outcome has since
been generalized for a plethora of asymptotically AdS theories
(for instance, \cite{cai}), including dynamical-boundary  
scenarios (for instance, \cite{sav}).\footnote{For a  thorough list of many
of the  relevant citations, see Ref.\cite{odie}.} 
Furthermore, the Cardy-Verlinde formula has even been extended 
to asymptotically de Sitter (dS) spacetimes 
(for instance, \cite{dan}\footnote{For further citations along this line,
consult Ref.\cite{med}.}),  although with only qualified success.\footnote{In
this regard, it is worth noting that the (proposed) dS/CFT correspondence
\cite{str2} remains on somewhat weaker ground than its AdS analogue.
This can especially be attributed to the
 lack of a globally timelike Killing vector
and the lack of a string-theoretical description. For further  references
on the dS/CFT duality, see  Ref.\cite{med}.}
\par
Just like the microscopic entropy calculations of (for instance) 
Carlip and Solodukhin \cite{carlip,sol},
the success of the Cardy-Verlinde formula (and its myriad of generalizations)
leads to a puzzling issue. Namely,  why can the degrees of freedom for
a  $d$-dimensional
field theory (albeit a conformal one) be consistently  explained in terms of
two-dimensional conformal formalism?
This will undoubtedly be  a difficult question to answer, given the
current absence of any reliable theory of quantum  gravity.
Very recently, however, Halyo did manage to shed some light on
the matter \cite{xhal}. In the context of a $d$+1-dimensional asymptotically
dS bulk, this author demonstrated that a dual ($d$-dimensional) CFT   
exhibits string-like behavior. In particular, the   CFT (or dS-bulk) entropy
was shown to be linearly related to the extensive energy
of the boundary theory.  On the basis of this result, Halyo
identified an effective string tension for the CFT.
With this  stringy description,
it follows that the $d$-dimensional boundary theory can 
alternatively be viewed as having only two relevant dimensions.
\par
Halyo went on to show \cite{xhal} 
that the CFT thermodynamic properties (including
the effective string tension) are directly related to those of a
string that lives on the ``stretched'' (cosmological) 
horizon\footnote{A stretched black hole (or
cosmological) horizon \cite{xsus,xstu} 
 refers to a timelike surface that lies at a distance of about
the fundamental string scale ($l_s$) above the event horizon.}
of the asymptotically dS bulk. (This latter picture follows
from a black hole-string correspondence that was originally
proposed by Susskind \cite{xsus}.) Moreover, when the thermodynamic properties
of the stretched  horizon have  properly been renormalized (see below), 
then the two stringy descriptions
turn out  to be essentially  equivalent.  
That is, the CFT can, at least effectively,
 be interpreted as a horizon-based string with a renormalized tension.
Within this inherently two-dimensional framework,  a Cardy-like
formulation of the CFT entropy begins to make sense. 
\par
Before proceeding to an outline of the current paper, a brief
discussion on the black hole-string correspondence is in order.
Roughly a decade ago, Susskind  conjectured \cite{xsus} a  one-to-one
correspondence between sufficiently massive
 black holes and highly excited fundamental strings.
This notion was based on the observation \cite{xtho} that, 
as string coupling increases, the size of a string state
must eventually become less than its Schwarzschild radius and,
hence, the string state must evolve into a black hole. Conversely,
as string coupling decreases, the size of a black hole will
fall below the string scale and, hence, the near-horizon geometry
can no longer be interpreted as a black hole. It follows that 
a black hole and some sort of string configuration must be strongly
and weakly coupled manifestations of the same entity.  Now consider
that, at large
enough mass, a typical string state consists of a small
number of highly excited strings. However, as such a  configuration
approaches the high-temperature conditions of the  horizon,
the state of a single string will become entropically preferred.
Alternatively,  an external observer might say that the strings have melted
together into a single string that ultimately fills up the
stretched horizon \cite{sss}.  Given these arguments, 
there should indeed be a one-to-one 
correspondence between black holes and strings.
\par
Contrary to the above philosophy, there is, of course, an observed discrepancy
between black hole and string thermodynamics. 
In particular, the entropy of a string has a linear energy dependence
while, for instance, the entropy of a  four-dimensional 
Schwarzschild black hole depends quadratically on its mass.
 However, Susskind proposed \cite{xsus}  
 that this discrepancy
can be accounted for when the energy of the string has  properly been
renormalized.  Such a renormalization follows naturally by virtue of
a large gravitational red shift that occurs between the stretched horizon
and an external observer. Exploiting the Rindler-like 
 near-horizon geometry (of a Schwarzschild black hole),
Susskind went on to demonstrate the feasibility of this proposal.
\par
The validity of this  black hole-string  correspondence  has 
since been rigorously substantiated  for a
wide range of black hole (as well as  brane) scenarios 
\cite{xhrs,xhkrs,xhp,xhal2,xdv,xdam,xugk}.  Generally speaking,
the  entropy  obtained from the counting of string states has been shown 
to describe the anticipated Bekenstein-Hawking  form up to some
ambiguity in the numerical coefficient.  It is worth noting 
that the same  philosophy has been applied to special
classes of supersymmetric extremal and near-extremal black holes.
(See, for instance, Ref.\cite{vaf} and, for a review,  Ref.\cite{xpeet}.)
For these calculations, the Bekenstein-Hawking form
has been reproduced exactly. However, a special property
of such  models (namely, supersymmetry protects the mass
against renormalization \cite{xsen}) prevents a more general application
of these particular methods.
\par
Now that the necessary background has (hopefully) been covered,
let us focus on the purpose of the current  study.  Here,
we will essentially be extending  Halyo's priorly discussed treatment 
\cite{xhal} to the case of an asymptotically AdS bulk (of arbitrary
dimensionality). That is, we will demonstrate that an appropriately  dual
CFT  can  effectively be described by a string
that lives on a  stretched black hole horizon. Although this is, perhaps,
a trivial extension, we believe that it is important to clarify
the success (or lack thereof) of this procedure for an AdS-bulk scenario. 
Our rationale being as follows. If any  progress is to be
made in extrapolating this outcome towards a microscopic realization
 of black hole entropy,
 then it will likely  come in an AdS (rather than dS) 
context,  where the holographic duality is much better understood. 
\par
The remainder of this paper is organized as follows. In Section 2,
we begin by introducing the bulk solutions of interest; 
namely, arbitrary-dimensional
AdS-Schwarzschild black holes and their topological variants \cite{bir}.
Next, we  derive the relevant black hole thermodynamics and,
 following Verlinde \cite{ver}, obtain the 
 corresponding expressions for a dually related CFT.
We  then  demonstrate the anticipated string-like behavior
of the CFT \cite{xhal} and accordingly identify an effective string tension
and energy.
\par
In Section 3, we consider a  boundary theory of a rather different nature;  
 the boundary now being the  AdS-black hole horizon.
Following  Susskind's proposal of a black hole-string 
correspondence \cite{xsus}, we show that the near-horizon 
geometry can  effectively be described by a  single fundamental
string that lives on the stretched horizon. In this process, 
the associated thermodynamics (including the string tension) are
unambiguously identified.
\par
In Section 4, we ultimately  show that these two  stringy
descriptions of the AdS-bulk are, indeed,  thermodynamically equivalent.
This entails a pair of coordinate rescalings  with respect
to the thermodynamic properties  of the stretched horizon. The first
rescaling is essentially a renormalization that accounts
for the gravitational red shift occurring between the horizon
and a  cosmological observer. The second rescaling is
necessary so that our hypothetical observer is located
precisely at the CFT boundary.  
\par
Finally, Section 5 contains
a summary and some commentary.

\section{A CFT Description of AdS}  

Let us begin by considering the model of interest. Namely,
Schwarzschild-like black hole solutions  in an
 $n$+2-dimensional anti-de Sitter background 
(i.e., $D=n+2$ Einstein gravity with a negative
cosmological constant).  In a suitably static gauge,  such
asymptotically AdS
  solutions
can be described by the following  line element \cite{bir}: 
\be
ds^2_{n+2}=-h(r)dt^2+{1\over h(r)}dr^2+r^2d\Omega^2_{n},
\label{1}
\ee
where:
\be
h(r)=k+{r^2\over L^2}-{\omega_{n}M\over r^{n-1}},
\label{2}
\ee
\be
\omega_{n}={16\pi G \over n \V}.
\label{3}
\ee
Here, $L$ is the curvature radius of the   AdS background
($L^{-2}=-2\Lambda/n(n+1)$, with $\Lambda$ as
 the negative cosmological constant),
$d\Omega_n^2$
denotes the line element of an $n$-dimensional
 hypersurface with a constant  curvature  $kn(n-1)$ and a volume
 $\V$, $G$ is
the $n$+2-dimensional Newton constant, and $M$ and $k$ are 
 constants of integration. $M$  is directly related to
 the  ADM (or conserved) 
 mass of the associated  black hole, which we will assume is always 
non-negative.\footnote{Although, technically speaking, black hole solutions
having  negative mass are allowable in the case of a hyperbolic
horizon geometry (i.e., the case of negative $k$) \cite{bir,emp}.}
Meanwhile, $k$ describes the horizon geometry and, without loss
of generality, can be set equal  to 
+1,0 or -1.  These three choices correspond to a horizon geometry
that is respectively spherical, flat or  hyperbolic.
\par
For a non-vanishing (positive) $M$, there will be a single black hole
horizon, which corresponds to the  positive root of $h(r)$.
Denoting this horizon by $r=R$, we thus obtain the following
useful relation:
\be
k+{R^2\over L^2}-{\omega_{n}M\over R^{n-1}}=0.
\label{4}
\ee
\par
The associated black hole thermodynamics  can  readily be identified.
For instance, if we disregard the energy of the $M=0$ vacuum  spacetime
(which is non-vanishing if $n$ is odd), then the excitation
energy of the black hole is simply given as follows \cite{bal}:
\be
E_{AdS}=M={1\over \omega_{n}}\left[{R^{n+1}\over L^2}+kR^{n-1}\right].
\label{5}
\ee
Furthermore,  the Hawking temperature \cite{hawt}  
can be obtained  with the usual
prescription: the inverse temperature  is equivalent to
the  periodicity of Euclidean time \cite{gh2}. This identification
yields:
\be
T_{AdS}={1\over 4\pi} \left.{dh\over dr}\right|_{r=R}={R\over 4\pi L^2}
\left[(n+1)+k(n-1){L^2\over R^2}\right].
\label{6}
\ee
Finally, the black hole entropy can be expressed in terms of
its ``area'' (i.e., the $n$-dimensional volume of the  horizon hypersurface)
via the Bekenstein-Hawking  definition \cite{bek,haw}:
\be
S_{AdS}= {\V R^n\over 4 G}={4\pi\over n\omega_n}R^n.
\label{7}
\ee
\par
Let us now  reconsider the thermodynamics from a new perspective:
an $n$+1-dimensional conformal field theory living on a timelike boundary of
the bulk spacetime.  In view of the well-established AdS/CFT correspondence 
\cite{mal,gub,wit}, such a boundary theory can  capably  provide
a holographic description of the  asymptotically AdS geometry.
In particular, the CFT thermodynamic quantities should be  identifiable
with those of the bulk black hole up to a simple red-shift factor.
Because of the conformal symmetry of the dual theory,
this red shift  can vary depending on  the choice of boundary
coordinates. Here, we follow Verlinde 
\cite{ver,sav}\footnote{Verlinde's analysis
was for $k=+1$ only. The $k\neq 0$
case was first considered by Cai \cite{cai}.} and fix the scale by 
setting the radial distance of the boundary equal to $R$ (i.e., equal to
 the radius
of the black hole horizon). This choice requires the bulk time to be rescaled
(on the boundary) such that $t\rightarrow tR/L$,
 which directly  translates into a  corresponding  red-shift factor of  $L/R$.
\par
On the basis of the above discussion, the CFT thermodynamics
can be expressed as follows:   
\be
E_{CFT}= {L\over R}E_{AdS}= {1\over \omega_{n}}\left[{R^n\over L}+
kL R^{n-2}\right],
\label{8}
\ee
\be
T_{CFT}={L\over R}T_{AdS}=
{1\over 4\pi L}
\left[(n+1)+k(n-1){L^2\over R^2}\right],
\label{9}
\ee
\be
S_{CFT}=S_{AdS}={4\pi\over n\omega_n}R^n. 
\label{10}
\ee
Note that the   entropy of the boundary theory is  never affected by the 
choice of scale 
\cite{wit}, as one might anticipate from the first law of CFT thermodynamics.
\par
Again following Verlinde \cite{ver},
we will  view the  CFT energy  as having arisen  from a pair 
of separable contributions. More specifically, 
$E_{CFT}=E_E+E_C$,  such that:  
\be
E_E= {1\over \omega_{n}}{R^n\over L},
\label{10.3}
\ee
\be
E_C={1\over \omega_{n}}
kL R^{n-2}.
\label{10.6}
\ee
Significantly to this split,
$E_E$ can readily  be identified as the  extensive contribution  to the 
total energy.  Hence, the remaining portion, $E_C$, is the sub-extensive
 contribution, which  is also known as the Casimir energy.\footnote{Note
that there is a factor of two  difference between this definition of $E_C$
and that of Verlinde's study \cite{ver}.}
\par
With these definitions, it is not difficult to confirm a  Cardy-like
form for the entropy \cite{car}; that is, the renowned Cardy-Verlinde
formula \cite{ver} (slightly modified in allowing for $k\neq +1$):
\be
S_{CFT}={4\pi\over n}R\sqrt{{E_C\over k}\left[E_{CFT}-E_C\right]}.
\label{cv}
\ee
Note that $E_C/k$ is to be regarded as a positive, non-vanishing quantity.
\par
It is also straightforward to verify the following form:
\be
S_{CFT}={4\pi L\over n}E_E.
\label{11}
\ee
\par
The significance of the above expression is the direct proportionality
that exists between the CFT entropy and a well-defined energetic
quantity, $E_E$.
Such  linearity is a characteristic behavior  of strings \cite{xpeet}, 
rather than black
holes (for which entropy  typically depends on  mass as $M^p$, 
with   $p>1$). With this in mind, we can re-interpret Eq.(\ref{11})
as the equation of state for an ``effective string'' having an energy: 
\be
{\cal E}_{CFT}={2C\over n}E_E
\label{11.6}
\ee
and having a tension:
\be
{\cal T}_{CFT}={C^2\over 2\pi L^2}.
\label{11.3}
\ee
Here,  $C$ is a yet-to-be-determined parameter that is possibly
a constant; however,
we anticipate  that, in general, $C=C(R;n,k,L)$ \cite{xhal}.
\par
The above string-like behavior is quite surprising, considering 
that the  boundary
theory has at least three spacetime dimensions (unless $n=1$).
On the other hand, there has been considerable evidence that
the relevant degrees of freedom for  {\it any} black
hole can be described by a two-dimensional conformal  theory 
\cite{carlip,sol}.\footnote{Also in support of this notion is 
the 
 Cardy-Verlinde description itself (\ref{cv}), given  
that the original Cardy formula \cite{car}
 has its genesis in two dimensions.}  It is possible that
 the above outcome  is yet another manifestation
of some  unknown, universal principle that seems to be  at work.

\section{A Stringy Description of AdS}

In the preceding  section, we ultimately demonstrated  that a 
dual CFT (with respect to an asymptotically AdS bulk)  exhibits an
 intriguing string-like behavior.  This observation segues nicely
into  our next topic; namely, a stringy description of   
asymptotically AdS spacetimes.  Here, we will be applying the
one-to-one correspondence
principle,   as  first advocated
by Susskind \cite{xsus},  between black holes and strings.
 The  underlying
premise is that a black hole and a fundamental string  can respectively
 be regarded   as strongly
and weakly coupled versions of the same entity \cite{xtho}. On this basis,
Susskind has proposed that a massive  black hole state can effectively be
described by  a highly excited string  which  (from the perspective
of a cosmological observer) fills up the so-called 
stretched horizon \cite{xstu,sss}.  Furthermore, Susskind
has argued that the obvious differences (between black hole  and string 
thermodynamics) can be attributed to a large gravitational
red shift that  occurs between the stretched horizon and an outside
observer. (For subsequent work along these lines, see Refs.\cite{xhrs,xhkrs,
xhp,xhal2,xdv,xdam,xugk}.)
\par
The Susskind program \cite{xsus} 
appears to apply quite  naturally given  any black hole
(or black object for that matter) with a   near-horizon geometry that
can  conform to a Rindler-like description.  Let us  now see how this
plays out for a black hole in a AdS background of arbitrary dimensionality.
\par
We begin here by reconsidering the asymptotically AdS geometry as
described by Eqs.(\ref{1}-\ref{3}). To study the near-horizon form,
let us introduce a new radial coordinate, $y$, in accordance with
$r=R+y$ (where $y<< R$ has been assumed). Up to first order in $y/R$,
the metric function $h(r)$ can now be written as: 
\be
h(y)\approx {\Delta (R)\over L} y, 
\label{12}
\ee
where we have defined:
\be
\Delta(R)\equiv (n+1){R\over L}+k(n-1){L\over R}.
\label{12.5}
\ee
\par
Applying the above, we find the following near-horizon form
for the  line element (\ref{1}):
\be
ds^2_{NH}=-y{\Delta(R)\over L}dt^2+
{L\over y\Delta(R)}dy^2+R^2d\Omega^2_{n}.
\label{13}
\ee
\par
It is useful if $y$ is replaced with a coordinate that
directly measures the proper  distance from any given point to  the horizon.
Denoting this proper distance as $\rho$, we have (up to the first perturbative
 order):
\be
\rho\approx \sqrt{{L\over \Delta}}\int^{y}{dy\over\sqrt{y}}
=2\sqrt{{L y\over \Delta }}.
\label{14}
\ee
The near-horizon line element (\ref{13})
now adopts the following Rindler-like form:
\be
ds^2_{NH}=-{\Delta^2 \over 4 L^2}\rho^2 dt^2+
d\rho^2+R^2d\Omega^2_{n}.
\label{15}
\ee
\par
To obtain a  near-horizon  geometry that is identically  Rindler
spacetime:
\be
ds^2_{NH}=-\rho^2 d\tau^2+
d\rho^2+R^2d\Omega^2_{n},
\label{17}
\ee
we simply redefine the time coordinate as follows:
\be
\tau\equiv {\Delta(R)\over 2L} t.
\label{16}
\ee
Hence, $\tau$ is the dimensionless Rindler time.
\par
Next, we consider thermodynamics as measured by a hypothetical
 Rindler observer at the stretched horizon. The dimensionless horizon
 temperature
is  known to be \cite{big}:
\be
T_R={1\over 2\pi}.
\label{18}
\ee
Meanwhile, the horizon entropy should still be given by the Bekenstein-Hawking
area law (or its $n$+2-dimensional analogue) \cite{bek,haw}. Hence:
\be
S_R=S_{AdS}.
\label{19}
\ee
\par
To determine the  dimensionless Rindler energy, we simply apply the first
law of thermodynamics; that is, $dE_{R}=T_{R}dS_{R}$.
This process yields:
\be
E_R={1\over 2\pi} S_{R}={2L\over n}E_E,
\label{20}
\ee
with the right-most relation following from Eqs.(\ref{10},\ref{11},\ref{19}).
\par
The linear relation between Rindler entropy and energy is (once
again) indicative of a string; in this case, one that lives on the stretched
horizon \cite{xsus}. Note that  the associated  string tension, in
dimensionless Rindler
coordinates, is trivially given by:
\be
{\cal T}_R={1\over 2\pi}.
\label{ten}
\ee
Of course, the dimensionless string energy is simply ${\cal E}_R=E_R$.
\par
Clearly, the fundamental length scale for 
 this effective
theory is just the string length, $l_s$ (as this determines the radial extent
 of the stretched horizon). Hence, we can obtain the
``true'' thermodynamics of the stretched horizon by 
 rescaling the dimensionless
Rindler quantities in terms  of  this length. That is, quantities
with units of energy (or inverse length) should be divided by $l_s$.
On this basis, the following identifications can readily be made:
\be
T_{SH}={T_R\over l_s}={1\over 2\pi l_s},
\label{22}
\ee
\be
S_{SH}=S_R= S_{AdS},
\label{huh}
\ee
\be
{\cal E}_{SH}= {{\cal E}_R\over l_s}={2L\over nl_s}E_E,
\label{23}
\ee
\be
{\cal T}_{SH}={{\cal T}_{R}\over l_s^2}={1\over 2\pi l_s^2}.
\label{21}
\ee
Note that, as expected, $T_{SH}$ corresponds precisely with the 
Hagedorn temperature
of a string \cite{xpeet}.
\par
We again point out that, according  to Susskind \cite{xsus},
the discrepancy between the above thermodynamic quantities
and  those of the corresponding black hole (in this case,
Schwarzschild-AdS or a topological variant) should
be attributable to the effects of an immense gravitational
red shift. This conjecture has, in fact,  been rigorously demonstrated for
a number of models (for instance, \cite{xhkrs}) up
to some ambiguity in the  numerical coefficients. Alternatively,
we will, in the next section, indirectly   demonstrate   the validity of 
this conjecture
(in an AdS-black hole context) by comparing the thermodynamic properties
for a pair of  stringy descriptions. These descriptions being the 
the apparent string living on
 the stretched horizon and
the effective string located at the CFT boundary.

\section{A String/CFT Correspondence?}

So far,  we have seen two different
 descriptions of an asymptotically AdS spacetime
that are indicative of a string living on a boundary; with the boundaries
in question being that which hosts a dual CFT and the stretched black
hole horizon. The purpose of the current section is to demonstrate
the equivalence of these  effective theories, with thermodynamics
serving as the testing ground.  Keep in mind that there is no
reason, {\it a priori}, to believe that these seemingly unrelated
pictures  should fundamentally coincide.
\par
As an initial step, it is necessary that the thermodynamics
of the stretched horizon be  rescaled for an  external or
cosmological
observer (who is presumed to be far away  from the horizon
in terms of all relevant length scales).
The simplest way to accomplish this task is to rescale the  Rindler
thermodynamic quantities (\ref{18}-\ref{ten}) so that they are
directly measured in terms of the ``cosmological'' time coordinate, $t$.  
Utilizing  Eq.(\ref{16}) (which directly relates $t$ to the Rindler
time, $\tau$), we  obtain the following rescaled  quantities: 
\be
T_{SH}^{\prime}={d\tau\over dt}T_{R}={\Delta(R)\over 4\pi L},
\label{24}
\ee
\be
S_{SH}^{\prime}=S_{R}=S_{AdS},
\label{26}
\ee
\be 
{\cal E}_{SH}^{\prime}={d\tau\over dt}{\cal E}_{R}={\Delta(R)\over n} E_E,
\label{25}
\ee
\be
{\cal T}_{SH}^{\prime}=
\left({d\tau\over dt}\right)^2 {\cal T}_{R}={\Delta^2(R)
\over 8\pi L^2}.
\label{27}
\ee 
Note that  a prime indicates that a rescaling has taken place. 
\par
It is worthwhile for us to compare the above outcomes
with those of Eqs.(\ref{22}-\ref{21}); that is, the
thermodynamic properties of the stretched horizon as
measured by a local observer.
In going from the horizon perspective to the cosmological one,
we observe  an  effective renormalization of 
roughly $\l_s^{-1}\rightarrow L^{-1}$.
Since it  is usually  assumed that $L>>l_{s}$, this translates
to a significant reduction in the energy, string tension and
temperature as measured by a distant observer.  Such
a renormalization is, however, expected
and    can be attributed to a large gravitational red shift
that naturally occurs between the horizon and an external 
vantage point \cite{xsus}. 
\par
Next, for the sake of comparison, 
we will consider an observer who is specifically  located at the CFT boundary.
Thus, to maintain consistency, it is  necessary 
that the coordinates be  rescaled  
so that the observer in question lives
on a timelike hypersurface of  fixed radial distance $R$ (i.e., the 
radius of the black
hole horizon). As previously discussed, 
this translates into  a red-shift factor
of $L/R$ for  any quantity having units of inverse length.
\par
Let us begin here with the stretched-horizon entropy. 
As usual, this dimensionless quantity
is unaffected by any coordinate rescalings, and so we trivially achieve the
following agreement:
\bea
S_{SH}^{\prime\prime}=S_{SH}^{\prime}&=& S_{AdS}
\nonumber \\
&=& S_{CFT}.
\label{29}
\eea
\par
Next, let us  consider the temperature at the stretched horizon.
The appropriate red shifting yields   the following:
\bea
T_{SH}^{\prime\prime}={L\over R}T_{SH}^{\prime} &=& {\Delta (R)\over 4\pi R}
\nonumber \\
 &=& {1\over 4\pi L}
\left[(n+1)+k(n-1){L^2\over R^2}\right].
\label{28}
\eea
Comparing with Eq.(\ref{9}), we  are able to make
 the anticipated  identification: $T_{SH}^{\prime\prime}=T_{CFT}$.
\par
Also on the agenda is the rescaled value of the string tension.
Here, we find:
\bea
{\cal T}_{SH}^{\prime\prime}={L^2\over R^2}{\cal T}_{SH}^{\prime} 
&=& {\Delta^2 (R)\over 8\pi R^2}
\nonumber \\
 &=& {1\over 8\pi L^2}
\left[(n+1)+k(n-1){L^2\over R^2}\right]^2.
\label{30}
\eea
Let us assume that this recaled string tension matches up
 with the CFT-inspired  
tension of Eq.(\ref{11.3}). (The validity of this assumption
will be tested via the energy.) 
This allows us to fix the parameter $C=C(R)$, and this process yields:
\be
C(R)={1\over 2}\left[(n+1) + k(n-1){L^2\over R^2}\right]
= {L\Delta(R)\over 2 R}.
\label{31}
\ee
\par
Finally, let us consider the rescaled value of the string energy.
This calculation goes as follows: 
\bea
{\cal E}_{SH}^{\prime\prime}={L\over R}
{\cal E}_{SH}^{\prime} &=& {L\Delta(R)\over nR}E_{E}
\nonumber \\
&=& {2C(R)\over n}E_E,
\label{32}
\eea
where we have incorporated  the identification of $C(R)$ into the lower line.
Remarkably, this outcome is in complete agreement with the CFT-inspired 
string energy, ${\cal E}_{CFT}$ 
of Eq.(\ref{11.6}). 
\par
As an aside, it  is interesting to
consider the special case of $n=1$ (i.e., three-dimensional AdS spacetime),
which (if $k=1$) describes 
 a theory that admits  BTZ black hole solutions \cite{btz}.
In this event, $C$ takes on the simple value
of $1$  and, hence, ${\cal E}_{SH}^{\prime\prime}={\cal E}_{CFT}=2E_{E}$. 
\par
Clearly, we have observed a precise correspondence between the
renormalized  thermodynamics of  the stretched horizon and the
thermodynamics of  the  conformal boundary theory. 
This apparent duality suggests
that the string which  lives on the CFT boundary 
(cf. Eqs.(\ref{11.6},\ref{11.3})) is
really just the string at the stretched horizon as viewed
by a CFT-based  observer. This is a remarkable outcome, considering
that we have utilized a pair of  descriptions with {\bf no}
 {\it a priori} relationship. 
\par
In spite of the success of this program, one might argue that
 the CFT ``total'' energy  (i.e., $E_{CFT}$ of
Eq.(\ref{8})), rather than the CFT-inspired string energy
(${\cal E}_{CFT}$),  should have been used to test  the duality.
If this is an accurate assessment, then there is an apparent failing in
the correspondence, inasmuch as   ${\cal E}_{SH}^{\prime\prime}=E_{CFT}$
can clearly {\bf not} be satisfied (except for  the very special
case of $M=0$ and  $k=1$; i.e., pure AdS).  
On the other hand,  such a discrepancy  seems very similar
to the difference in energy when one compares calculations
in, for instance, global and Poincare AdS coordinates \cite{bal}.
That is to say, conserved charges in an AdS  background
are not  so well defined (as they are in asymptotically flat
spacetimes) and may indeed be subject to an ambiguous
observer dependence. 
\par
Regardless of the viewpoint one takes on the above matter, 
it is still an interesting exercise to 
calculate the explicit difference between  ${\cal E}_{SH}^{\prime\prime}$
and $E_{CFT}$. 
First, let us apply Eqs.(\ref{31},\ref{32}) and re-express
 the renormalized energy of the stretched horizon
as follows:
\be
{\cal E}_{SH}^{\prime\prime}
={R^n\over n L\omega_n}\left[ (n+1) +k(n-1){L^2\over R^2}\right].
\label{33}
\ee
Also, let us recall Eq.(\ref{8}) for the total  energy of the CFT:
\be
E_{CFT}= {R^n\over n L\omega_n}\left[ n +kn{L^2\over R^2}
\right].
\label{34}
\ee
Defining
 $\delta E \equiv {\cal E}_{SH}^{\prime\prime}-E_{CFT}$,  we then have:
\be
\delta E= {R^n\over n L\omega_n}\left[1 -k{L^2\over R^2}
\right] ={1\over n}\left[E_E-E_C\right],
\label{35}
\ee
where Eqs.(\ref{10.3},\ref{10.6}) have also been  incorporated.
\par
The simplicity of the above expression is quite intriguing. We suspect
that there is some profound explanation as to why the ``energy
shift'' takes on such a concise, universal form; however, the answer remains
a mystery at the present time.

\section{Conclusion}

In summary, we have been studying an  asymptotically anti-de Sitter
 spacetime of arbitrary dimensionality. In particular, we have  concentrated
on static solutions having a single black hole horizon; that is,
 Schwarzschild-AdS black holes and  their
topological variants  \cite{bir}.  To begin the analysis, we  introduced
the relevant formalism and then derived the black hole
thermodynamic expressions for this somewhat generic model. 
Following these preliminary  
considerations, we proceeded to investigate  a pair of effective
boundary theories that are related to the bulk  theory
via commonly accepted  dualities. (We once again note that this
work has been  inspired by  Halyo's similar treatment in a de Sitter
context \cite{xhal}.) 
\par
The first boundary theory under consideration was   a conformal
field
theory that lives on a timelike  hypersurface  of the AdS bulk.
Exploiting the AdS/CFT duality \cite{mal,gub,wit}, we
were able to identify the  corresponding thermodynamic properties. 
 In particular, we verified the celebrated Cardy-Verlinde
formula \cite{ver}  and, moreover, demonstrated a linear 
relation  between the entropy and
 extensive energy of the CFT. 
Significantly, this  linear relationship
 indicates  that the CFT can effectively be  viewed
as a  string that lives on this cosmological boundary. On this basis, we were
able to identify  an effective  tension and energy 
   for this
 stringy version of the conformal boundary theory.
\par
The second  boundary  theory of interest was based on
Susskind's proposed correspondence \cite{xsus} between any 
sufficiently massive black 
hole    and a highly excited string that fills up the  stretched
horizon. This  duality becomes evident when one
transforms the near-horizon geometry of a black hole into a Rindler-like
form.  Following this program, we are able to calculate
the thermodynamic properties of the (conjectured) string at the
 stretched horizon (which follow directly from
the thermodynamics of the associated Rindler space). The black hole-string
correspondence \cite{xsus} 
then implies that a large gravitational red shift should
account for the thermodynamic differences  between 
 this string state  and  the associated black hole.   
\par
The final portion of our analysis focused on comparing this
pair of (seemingly unrelated) boundary theories. Such a comparison
necessitated a two-step procedure.  Firstly, we  reconsidered
the stretched-horizon thermodynamics from the  perspective of
a bulk observer (far away from the horizon). This required a 
coordinate transformation from Rindler time to ``cosmological'' time,
for which the net effect was a significant renormalization
of the dimensional thermodynamic properties.  Such a  renormalization
(or red shifting)
is expected  by virtue of the long-range gravitational field
that intervenes between the stretched horizon and the external
observer. Secondly,  for a direct comparison of these effective
theories, it was necessary for our bulk observer to be located
at the CFT-based hypersurface.  This consideration was accounted for
with one additional red shifting of the stretched-horizon thermodynamics.
\par
Ultimately, we were able to show that the renormalized thermodynamics of
the stretched horizon are indeed equivalent to the ``stringy'' CFT 
thermodynamics (provided that an arbitrary parameter has been 
suitably fixed). These  identifications included the temperature,
string energy, string tension and, trivially, the entropy.
This remarkable equivalence implies that the thermodynamics on the
CFT boundary can directly
be attributed to  a string that
lives on the stretched horizon. This is a significant outcome given
that the two boundary theories are {\it a priori} unrelated.
\par
One issue of note is a finite  shift that occurred between the (renormalized)
string energy and the  ``standard energy''  which is usually
attributed to the CFT \cite{ver}.  We have interpreted this
discrepancy as being analogous to an  ambiguous observer dependence  
that is inherent to asymptotically
AdS spacetimes. That is, different AdS coordinate systems
can often  give rise to different definitions of energy (and other
conserved charges) \cite{bal}.  Interestingly, this energy shift
took on a  universal, simple form: essentially, the difference
between the extensive and sub-extensive (or Casimir) 
contributions to the CFT energy. We expect that there is some
profound explanation for this outcome and hope to return to
this issue in a future study.
\par
 In conclusion, it is quite puzzling that a  CFT
of arbitrary  dimensionality can be explained, at least
effectively, by a single  fundamental string; that is,  a
two-dimensional entity.  On the other hand, this outcome
nicely complements the success of the ($n$+1-dimensional) 
Cardy-Verlinde formula \cite{ver};
insofar as  the original Cardy formula should, in principle, only  apply to
two-dimensional CFTs \cite{car}.   The notion
that a black hole can holographically  be described by a two-dimensional
CFT is certainly not a new one \cite{crev}. In fact, Carlip \cite{carlip}
and Solodukhin \cite{sol} have demonstrated that this is
 truly  a universal feature of black hole geometries.  The pertinent point is
that, near a black hole horizon, the relevant physics appears to be 
restricted to the $r$-$t$ plane. It is remarkable that
so many treatments, which are otherwise unrelated, can reach the
same conclusion.
(What is,  perhaps, even more
remarkable is the  apparent encoding of this near-horizon information  on
a distant boundary. This, in a nutshell, is the power
of the Holographic principle.)   
It seems most likely that these various viewpoints  represent
 different descriptions of one fundamental theory.
A  better  understanding of  this phenomena 
would  go a long way towards resolving   
the microscopic origin of black hole entropy.

\section{Acknowledgments}
\par
The author  would like to thank  V.P.  Frolov  for helpful
conversations.

\end{document}